# An experiment that proves the absence of nonlocality in quantum mechanics

C. S. Unnikrishnan

*Gravitation Group, Tata Institute of Fundamental Research, Homi Bhabha Road, Mumbai - 400 005, India* &

*NAPP Group, Indian Institute of Astrophysics, Bangalore - 560 034, India.*

E-mail addresses: unni@tifr.res.in

**I have been arguing that quantum nonlocality, deeply entrenched in the present formalism of quantum mechanics and widely believed as a reality by physicists, is in fact absent[1]. Spooky nonlocal state reduction is the most, and perhaps the only irrational feature of present day physics. There are experimental results that reject nonlocal state reduction at a distance[2]. Also, there are arguments that show that signal locality itself can be violated if there is true nonlocal collapse of the wavefunction. The Bell's inequalities – the violation of which polarized physicists in favour of nonlocality – arise not due to nonlocality, but due to ignoring prior information on correlations encoded in the phase of local probability amplitudes[1]. Here I discuss an experiment involving particles entangled in energy and time variables that shows that there is no nonlocal state reduction during measurements on entangled particles. Quantum mechanics is inconsistent if it includes the concept of wavefunction collapse as a physical process for entangled multi-particle systems.**

It has been pointed out several times that there are strong reasons from experimental observations as well as from consistency requirements that there is no quantum nonlocality and state reduction at a distance during measurements on entangled quantum systems[1,2]. But, quantum nonlocality is a deeply entrenched concept in the present day interpretation of quantum mechanics despite its ethereal qualities. It is therefore understandable that there is strong resistance to any denial of quantum nonlocality. However, unambiguous results from experiments, either gedanken or real, should convince physicists sooner or later that the concept of quantum nonlocality has to be abandoned. The reasons why the violation of Bell's inequalities does not imply nonlocality are discussed already[1]. Also, it has been shown that there is a class of local theories that reproduce quantum correlations, in which the prior correlations at source are encoded as a phase information in local amplitudes[1]. An experiment[3], with particles entangled in momentum and position that conclusively show that there is no nonlocal state reduction was discussed earlier[2]. Here I discuss another experiment with particles entangled in energy and time that proves that nonlocal state reduction is indeed absent in quantum phenomena as a physical process.

The two-particle entangled state we consider is very similar to the original entangled EPR state[4] $\delta(x_1 + x_2)$. When the entanglement is in energy, the state is $|1,2\rangle \sim \delta(E_1 + E_2 - E_0)$. Such states are produced in down conversion of laser light of energy $E_0$ into two photons of energies $E_1$ and $E_2$.

The basic experimental scheme is shown in Figure 1. A source produces the two-particle entangled state and the two particles are made to go in opposite directions. The physical process that generates the two-particle state happens in a time scale $\tau_s$. For the photon down-conversion process, this time scale is of the order of pico-seconds. If a measurement of time-of-arrival is made on the two photons, they will be detected within a temporal window of $\tau_s$. In the experiment, we gate the source with a time window $\tau_g$ such that $\tau_g \gg \tau_s$. Clearly this does not affect the distribution of difference in times of arrival for the two photons.

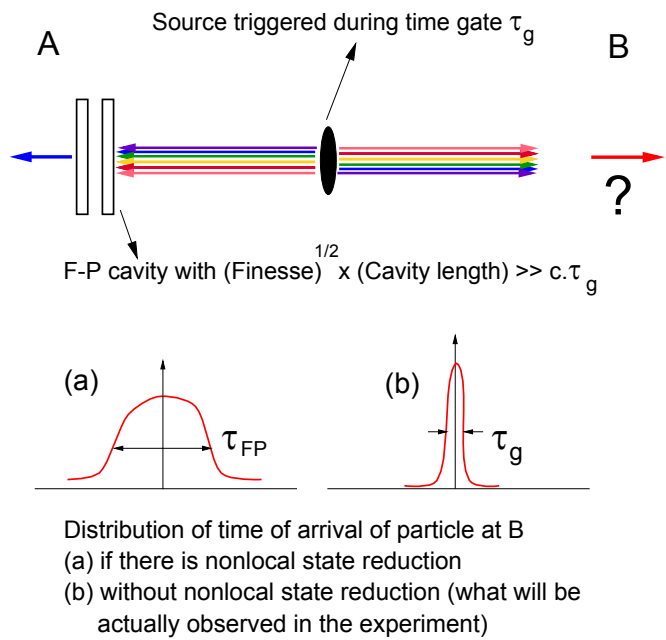

Figure 1: *The upper section shows the scheme of the experiment in which time of arrival of two entangled particles are measured with reference to a trigger generated from a pulsed source (The colors merely indicate the correlations that will be observed when a measurement is made, and not "what is" before a measurement). The lower part shows the drastic difference between (a) the prediction from quantum mechanics with nonlocal state reduction and (b) prediction for the experimental outcome from quantum mechanics without nonlocal state reduction.*

Now, one of the photons is sent to a high finesse Fabry-Perot filter (for particles this will be a narrow bandwidth energy analyzer). This will make the state of the first particle to be realized in a very narrow energy range. The finesse $F$ and the length $L$ of the F-P cavity is chosen such that $\tau_{FP} = F^{1/2}L/c \gg \tau_g$. This is equivalent to choosing the product of the Quality factor of the cavity and its length to be much larger than $\tau_g$. Then the fundamental principle of quantum mechanics also dictates that the time of arrival of the particle after the F-P filter will be considerably more spread out than the time scales $\tau_s$ or $\tau_g$. This spread will be approximately $\tau_{FP}$ and can be measured relative to the trigger from pulsing the source, with small error since $\tau_{FP} \gg \tau_g$. This is a manifestation of the physical fact that the state of the first particle is approximately $\delta(E_1)$, with an uncertainty of $\Delta E \approx \hbar/(F^{1/2}L/c)$.

What does instantaneous nonlocal collapse of the state of the second particle, as demanded by the standard interpretation, imply? Obviously, an instantaneous nonlocal collapse would mean that the second particle is in a state sharply peaked at energy $E_2$, immediately after the first particle has passed through the F-P cavity. Quantum theory predicts that the sharpness with which this state is defined after the nonlocal collapse is same as that of the first particle. Then the fundamental principle of quantum mechanics imply that the time of arrival of the second particle, relative to the trigger pulse from the source should be as spread out as that of the first particle. Otherwise, the second particle is not in a definite state of energy. This can easily be checked experimentally.

In fact, it will be found that the time of arrival of the second particle is not spread out to $\tau_{FP}$, and that it is still of the order of $\tau_g$. This conclusively rules out the hypothesis that the second particle's state collapsed to a state sharply peaked in energy around $E_2$ as a result of the measurement of energy on the first particle. There is no nonlocal state reduction during measurements on entangled quantum systems.

Every experiment that is trying to study the physical characteristics of nonlocal collapse, including the "speed of transmission of the nonlocal influence", is bound to fail since there is no nonlocal collapse. A measurement on one particle does nothing to the physical state of the space-like separated entangled particle. This result can be proved within the mathematical formalism of quantum mechanics[2] and can be checked experimentally in more than one experimental configurations.

**References:**

1. Unnikrishnan C. S., Is the quantum mechanical description of physical reality complete ? Resolution of the EPR puzzle, *Found. Phys. Lett.*, **15**, 1-25, (2002); Also see *quant-ph/0001112*; *quant-ph/0004089*; *Current Science*, **79**, 195, (2000); *Annales de la Fondation L. de Broglie,* **25**, 363 (2000).

2. Unnikrishnan. C. S., Einstein was right: Proof of absence of nonlocal state reduction in quantum mechanics, quant-ph/0206175, (2002).

3. Kim, Y-H., and Shih, Y., Experimental realization of Popper's Experiment: Violation of the Uncertainty Principle? *Found. Phys.*, **29**, 1849-1856, (1999).

4. Einstein, A., Podolsky, B., and Rosen, N., Can quantum mechanical description of physical reality be considered complete?, *Phys. Rev.* **47**, 777-780 (1935).

Acknowledgments: I thank R. Cowsik for several discussions. This version of the thought experiment owes much to my recent discussions with him on quantum nonlocality in interferometer experiments employing photons entangled in energy and time.